\documentclass[twocolumn,tighten]{aastex6} 
\pdfoutput=1 
\usepackage{amsmath,amstext}
\usepackage[figure,figure*]{hypcap}

\shorttitle{Polycrystalline crusts in accreting neutron stars}
\shortauthors{Caplan et al.}

\begin{document}

\title{Polycrystalline crusts in accreting neutron stars}

\author{M.~E.~Caplan\altaffilmark{1}, Andrew Cumming\altaffilmark{1}, D.~K.~Berry\altaffilmark{2}, C.~J.~Horowitz\altaffilmark{2}, R.~Mckinven\altaffilmark{3}}
\affil{
\altaffilmark{1}{Department of Physics and McGill Space Institute, McGill University, 3600 rue University, Montreal, QC, H3A 2T8, Canada}\\
\altaffilmark{2}{Department of Physics and Nuclear Theory Center, Indiana University, Bloomington, IN 47405, USA}\\
\altaffilmark{3}{Department of Astronomy \& Astrophysics and Dunlap Institute of Astronomy \& Astrophysics, University of Toronto, ON M5S 3H4, Canada}
}

\email{matthew.caplan@mcgill.ca}

\begin{abstract}
The crust of accreting neutron stars plays a central role in many different observational phenomena. 
In these stars, heavy elements produced by H-He burning in the rapid proton capture (rp-) process continually freeze to form new crust. In this paper, we explore the expected composition of the solid phase. We first demonstrate using molecular dynamics that two distinct types of chemical separation occur, depending on the composition of the rp-process ashes. We then calculate phase diagrams for three-component mixtures and use them to determine the allowed crust compositions. We show that, for the large range of atomic numbers produced in the rp-process ($Z\sim 10$--$50$), the solid that forms has only a small number of available compositions. We conclude that accreting neutron star crusts should be polycrystalline, with domains of distinct composition. Our results motivate further work on the size of the compositional domains, and have implications for crust physics and accreting neutron star phenomenology.
\end{abstract}


\section{Introduction}

The structure and composition of accreting neutron star crusts plays a key role in a diverse set of astrophysical phenomena. As the crust is compressed by accretion, nuclear reactions such as electron captures or pynonuclear fusion change the composition and deposit heat \citep{Haensel2003}. In transiently-accreting neutron stars, cooling of the hot crust during periods of quiescence has been used to infer properties such as the crust thermal conductivity \citep{Shternin2007,Brown2009} and to limit the heat capacity and neutrino emissivity of the neutron star core \citep{Heinke2010,Cumming2017,Brown2018}. The electrical conductivity sets the rate of decay of crust currents \citep{Urpin1995,Brown1998}, important because accreting neutron stars, as progenitors of millisecond radio pulsars, are believed to be undergoing accretion-driven magnetic field decay (e.g.~\citealt{Konar2017}). Depending on the shear modulus \citep{Horowitz:2007hx}, asymmetries in electron capture layers may give quadrupole moments large enough to limit the spin-up of the neutron star by angular momentum loss from gravitational radiation \citep{Bildsten1998}, potentially observable with next generation gravitational wave observatories \citep{Watts2008}.

\begin{figure*}[ht!]
\centering
\includegraphics[width=0.42\textwidth]{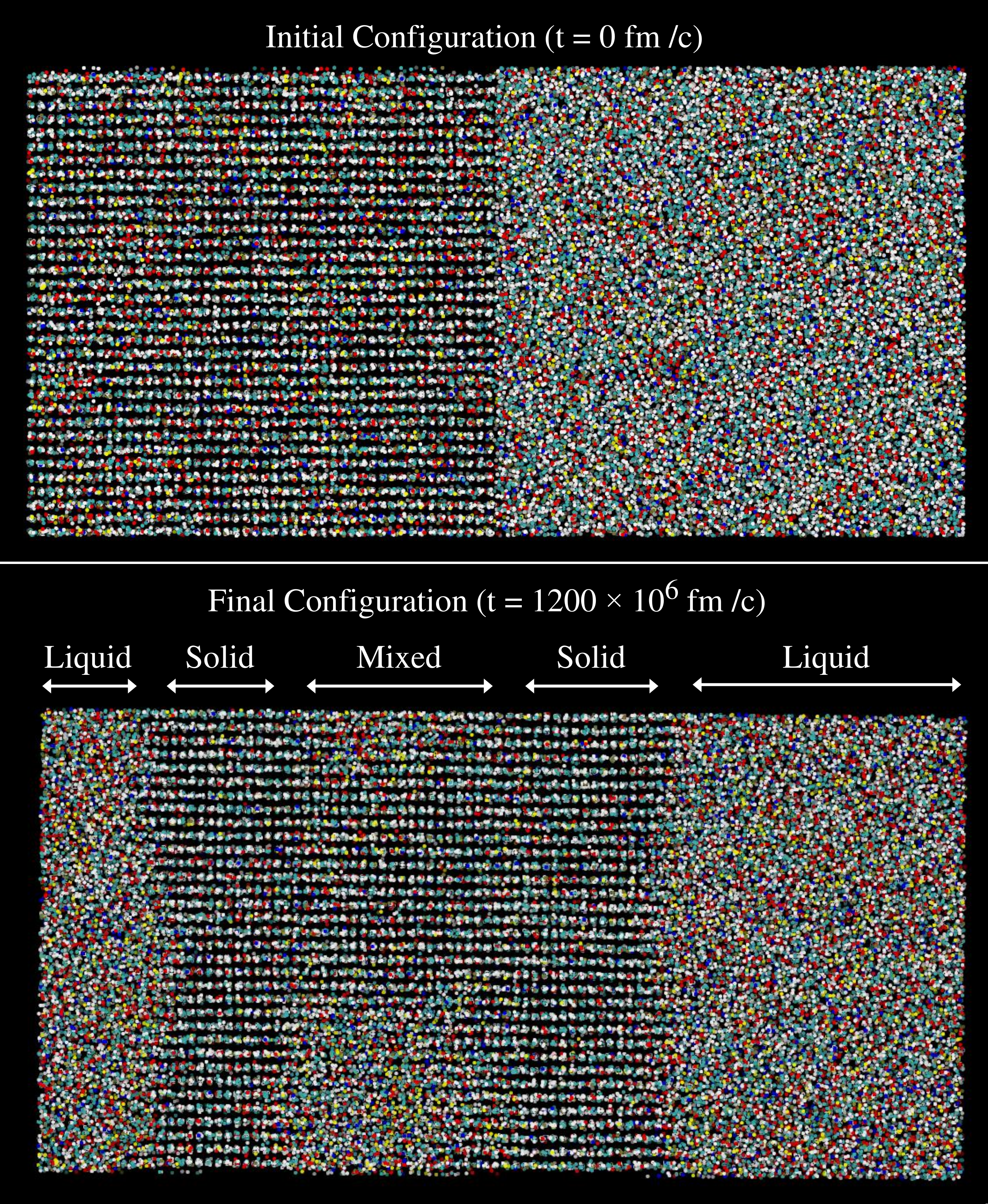}
\includegraphics[width=0.56\textwidth]{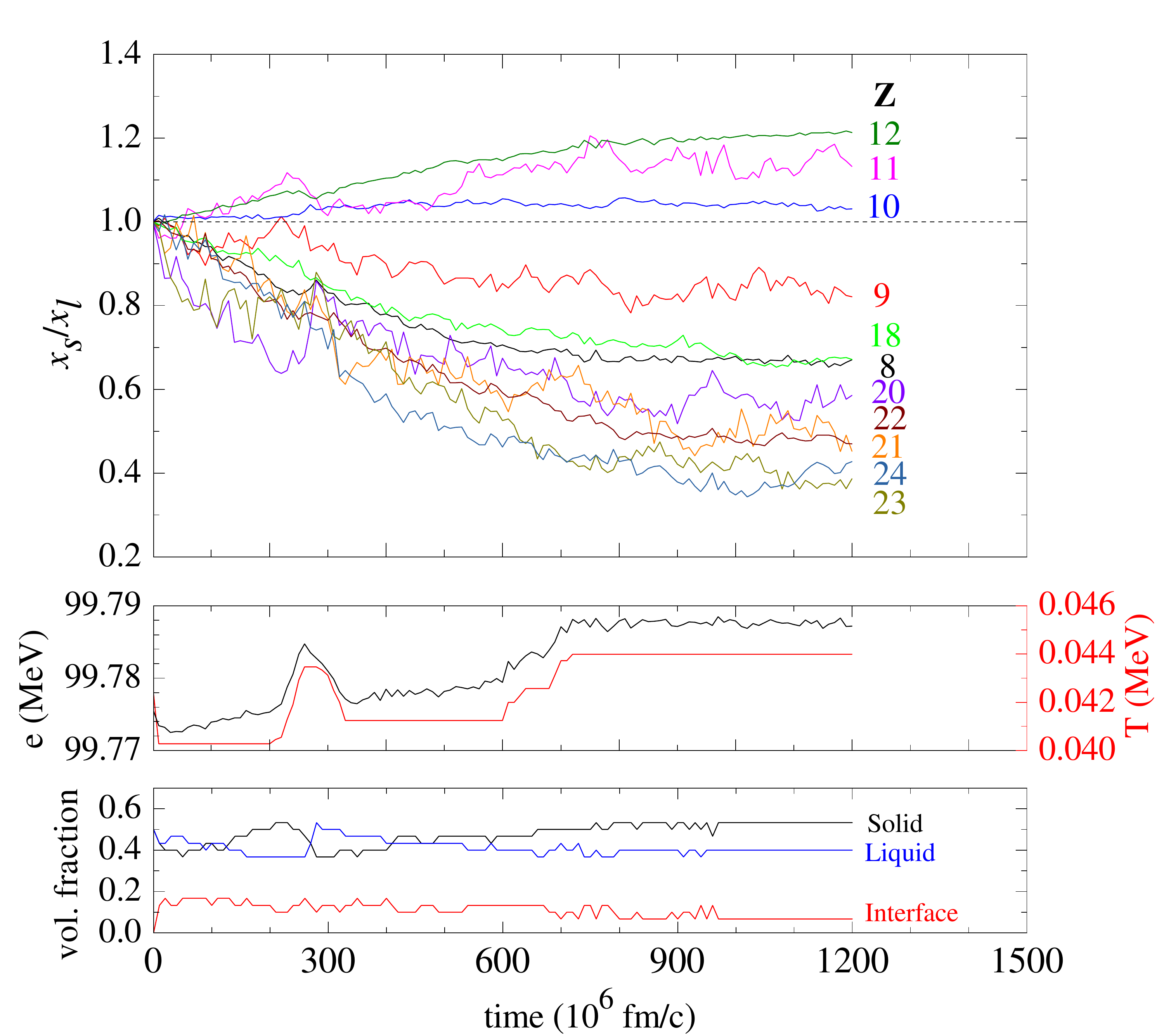}
\caption{\label{fig:MD} Molecular dynamics simulation. {\bf Left.} Initial and final configurations, with each point representing one nucleus. The simulation is 3D, but an orthographic projection is used here for clarity\added{, and shows the bcc (100) surface in the solid}. The clustering in the solid is due to the distribution of nuclei in the third dimension; the simulation is approximately a bcc lattice and each point is near a lattice site. The colors are dimmed according to depth in the field, and also identify ion species: in order of abundance they are $Z=12$ white, $Z=10$ cyan, $Z=8$ red, $Z=18$ blue, $Z=11$ tan, and the remaining six species (totaling 5\% of all ions) shown in yellow. 
During the simulation the liquid region has moved over the periodic boundary, and a subregion of liquid has developed within the solid region (see text for discussion). {\bf Right.} Thermodynamic history showing the evolution of abundances in the solid and liquid, thermal history (with energy per nucleus $e$), and volume fractions.} 
\end{figure*}

\begin{figure*}[ht!]
\centering
\includegraphics[width=0.9\textwidth]{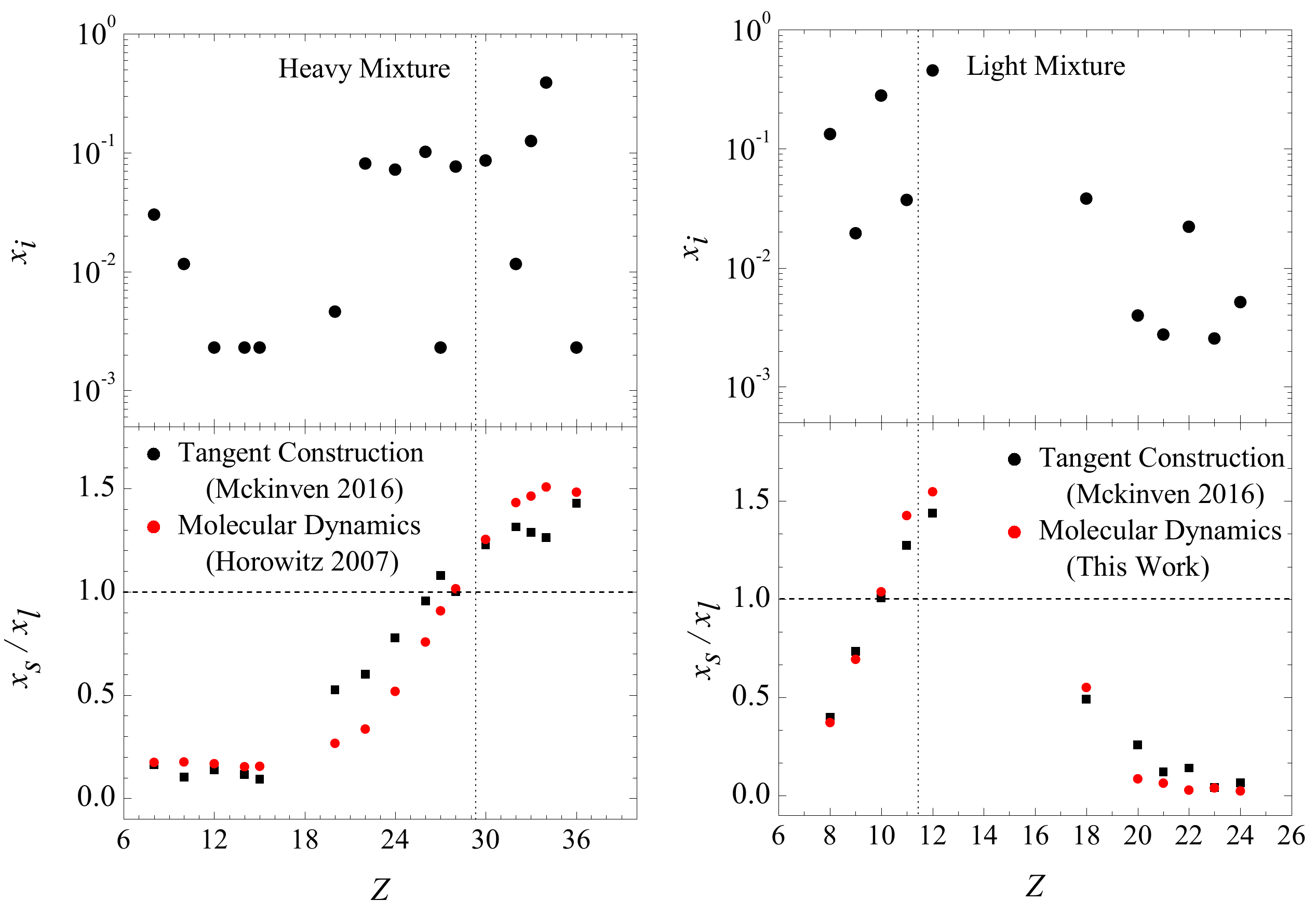}
\caption{Comparison of the phase separation of heavy and light mixtures. The top panels show the number abundance of each species produced in the rp-process. The bottom panels show the equilibrium solid-liquid abundance ratio calculated using the double tangent construction method \citep{Mckinven:2016zkg} (black) and molecular dynamics simulations (red) (taken from \citealt{Horowitz:2007hx} for the heavy mixture). The vertical dotted lines shows the mean charge for each initial mixture; the horizontal dashed line indicates equal abundances in the solid and liquid phase.
\label{fig:MD_comparison}}	
\end{figure*}

Because many accreting neutron stars have accreted enough matter ($\gtrsim 10^{-2}\ M_\odot$) to replace their entire crust, it is crucial to understand how freshly-accreted matter freezes to form new solid. 
Freezing is well-understood for a one-component plasma (OCP), which freezes into a bcc lattice when the density is high enough for the Coulomb interactions between nuclei to overcome their thermal motion (e.g.~\citealt{Brush1966,Potekhin2000}). The material freezing to make the new crust in accreting neutron stars, however, consists of a complex mixture of elements produced by nuclear burning of the accreted H and He via the rp-process \citep{Schatz1999}. Investigations of the lattice structure and freezing point of such mixtures have only started to be carried out more recently \citep{Horowitz:2007hx,Horowitz:2009af,HCB2009,Mckinven:2016zkg}.

An important process in multicomponent plasmas is chemical separation on freezing. In a molecular dynamics study of one realistic composition, \cite{Horowitz:2007hx} found that light nuclei (low atomic number $Z$) were retained in the liquid phase, with heavier nuclei (high $Z$) going into the solid. This change in composition raises the question of how freezing accommodates the large variety of different abundance patterns that are produced by the rp-process \citep{Schatz2003}. Recently, \cite{Mckinven:2016zkg} used a double-tangent-construction method based on analytic fits to free energies to calculate the liquid-solid equilibrium for a large sample of rp-process ashes. They identified a new type of chemical separation for light mixtures in which the solid forms primarily from the most abundant element, with other elements (light and heavy) going into the liquid phase. This opens a channel for light composition crust to be formed.

In this work, we explore the expected composition of accreting neutron star crusts. We first use molecular dynamics simulations to confirm that two possible phase separations -- light and heavy -- can occur (\S 2). We then survey the possible solids that can form, using a three-component approximation that allows a calculation of the full phase diagram. We show that, as a consequence of the large range of atomic number $Z$ in the mixture, only a small number of solid compositions are available to form new crust. For a given rp-process ash, the crust must therefore consist of distinct solid phases; accreting neutron stars have a polycrystalline crust whose domains have distinct compositions. In \S 4 we summarize our results and discuss the implications for accreting neutron stars.

\section{Molecular dynamics simulations of the phase separation of a light mixture}

In this section, we perform a large molecular dynamics simulation to determine the chemical separation on freezing for one of the light mixtures considered by \cite{Mckinven:2016zkg}. Their work was based on analytic expressions for the free energy of multicomponent plasmas extrapolated from simulations of two and three component plasmas \citep{Ogata1993,Medin2010}; here we compare their results with a direct molecular dynamics simulation.

Our molecular dynamics simulations treat nuclei as classical point particles with charge $Z_i$ interacting via a two-body potential, 
\begin{equation}
V(r_{ij})=\frac{Z_i Z_j e^2}{r_{ij}} \exp(-r_{ij}/\lambda),
\label{eq.V}
\end{equation}  
where $r_{ij}$ is the separation between two nuclei. The exponential screening, due to the degenerate electron gas between ions, is calculated from the Thomas Fermi screening length $\lambda^{-1}=2\alpha^{1/2}k_F/\pi^{1/2}$ where the electron Fermi momentum $k_F=(3\pi^2n_e)^{1/3}$ and $\alpha$ is the fine structure constant. The electron density $n_e$ is equal to the ion charge density, $n_e=\langle Z\rangle n$, where $n$ is the ion density and $\langle Z\rangle$ is the average charge. The screening length is generally greater than the inter-ion spacing due to the high electron Fermi energy in the neutron star crust. We solve Newton's equations of motion numerically to study the evolution of the material using the Indiana University Molecular Dynamics (IUMD) CUDA-Fortran code, version 6.3.1, which has previously been used to study neutron star crusts \citep{astromaterials}.

The composition we consider is taken from \cite{Schatz1999} (accretion rate $\dot{m}=0.1\,\dot{m}_{\rm Edd}$, helium fraction $Y=0.2752$), removing species less abundant than $10^{-3}$, and assuming that nuclei with $Z \leq 6$ have burned to Mg ($Z=12$) (the abundances are shown in the right panel of Fig.~\ref{fig:MD_comparison}). In reality, the burning produces many more different species than shown here. For simplicity, we include only the 11 most abundant elements. We take all isotopes of the same element to have the mass of the most abundant isotope.

Following the procedure of \cite{Horowitz:2007hx}, we first initialize a smaller simulation of 3,456 nuclei ($12 \times 12 \times 12$ bcc unit cells) with periodic boundary conditions and use trial and error to find the melting/freezing temperature. We set the density to $n = 7.18 \times 10^{-5}$ fm$^{-3}$ (screening length $\lambda = 35.8516\ {\rm fm}$), initial temperature $T_0=0.40\ {\rm MeV}$, and timestep $dt = 25\ {\rm fm/c}$. (Although these densities and temperatures are higher than typical conditions at the freezing depth in a neutron star crust, the evolution depends on $T$ and $\rho$ only through the Coulomb parameter $\Gamma = \langle Z^{5/3}\rangle e^2/a_ek_BT$ with $a_e=(3/4\pi n_e)^{1/3}$, which is chosen here to be near the freezing point). \replaced{The nuclei were initialized with random positions, heated to 0.130 MeV to equilibrate, and}{The nuclei were initialized with random positions and with velocities generated from a Maxwell-Boltzmann distribution at the desired temperature with zero net momentum. A trial-and-error process was used to find the freezing temperature over $2240 \times 10^6$ fm/c ($89.6 \times 10^6$ timsteps); first, the simulation was slowly heated to 0.130 MeV by periodically rescaling the velocities, and then } supercooled to 0.036 MeV where the mixture froze (molecular dynamics simulations tend to freeze nearer $0.9\ T_{\rm melt}$ due to finite time effects). \added{Coincidentally, when the mixture froze the crystal planes that formed were parallel with the boundaries simulation (seen in Fig. \ref{fig:MD}). This may be a consequence of simulating with $12^3 \times 2 $ nuclei which are able to perfectly fill a cubic simulation volume with a bcc lattice.} The crystal was then tiled to make a $2 \times 2 \times 2$ larger volume configuration with 27,648 nuclei, and then \replaced{equilibrated}{evolved} for an additional $400 \times 10^{6}\ {\rm fm/c}$ ($16 \times 10^{6}$ timesteps) and heated to $T=0.042\ {\rm MeV}$, which was believed to be nearer the true melting point. \added{This verified that the larger tiled crystal simulation was stable and would not spontaneously melt at the higher temperature.}

A liquid configuration with 27,648 nuclei and an identical composition was assembled in a similar way, by tiling a smaller liquid configuration containing 3,456 nuclei at $T = 0.0413$ MeV, then equilibrated for $50 \times 10^{6}$ fm/c ($2 \times 10^{6}$ timesteps) at $T=0.042\ {\rm MeV}$.

\begin{figure*}[t!]
\centering
\includegraphics[width=0.49\textwidth]{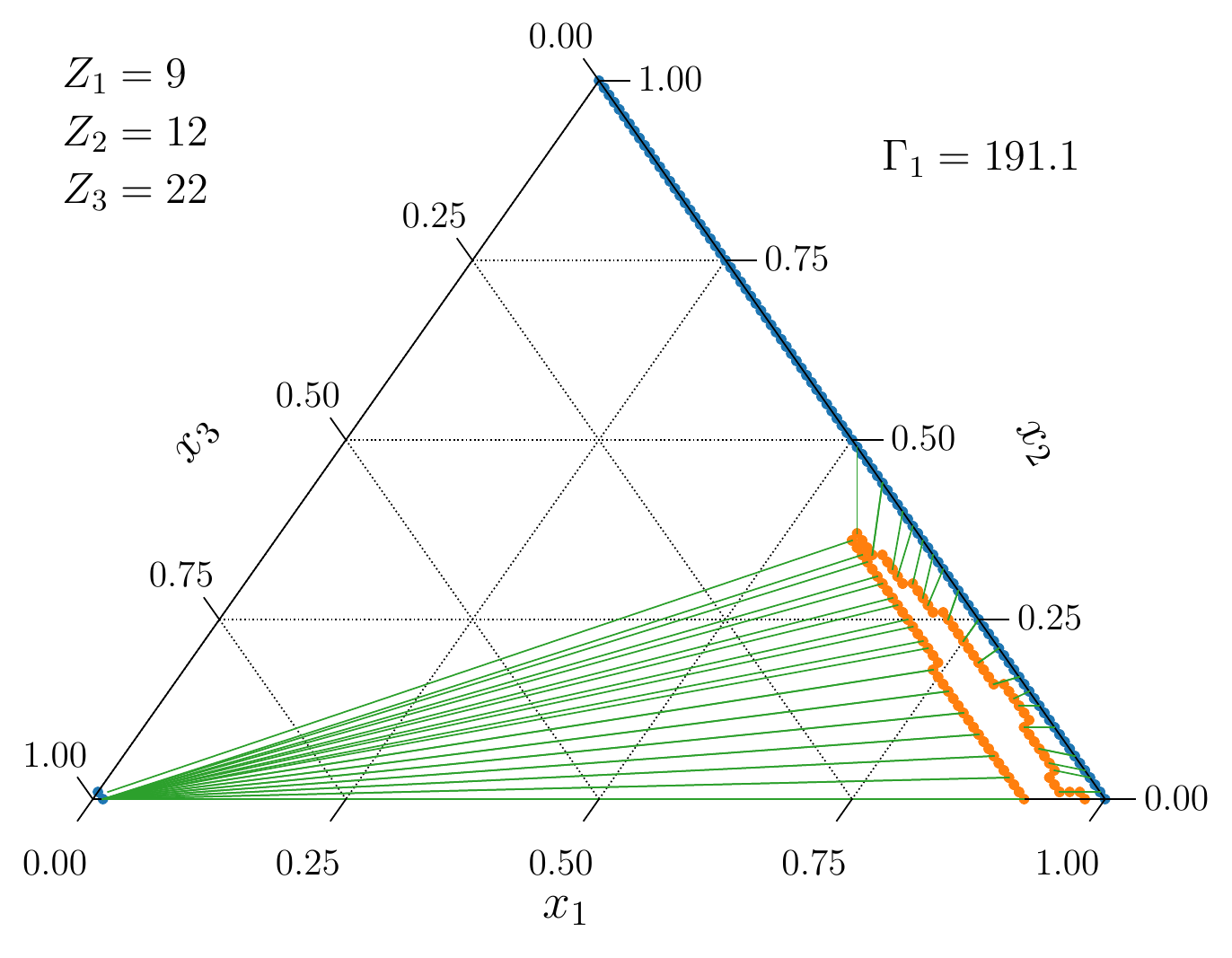}
\includegraphics[width=0.49\textwidth]{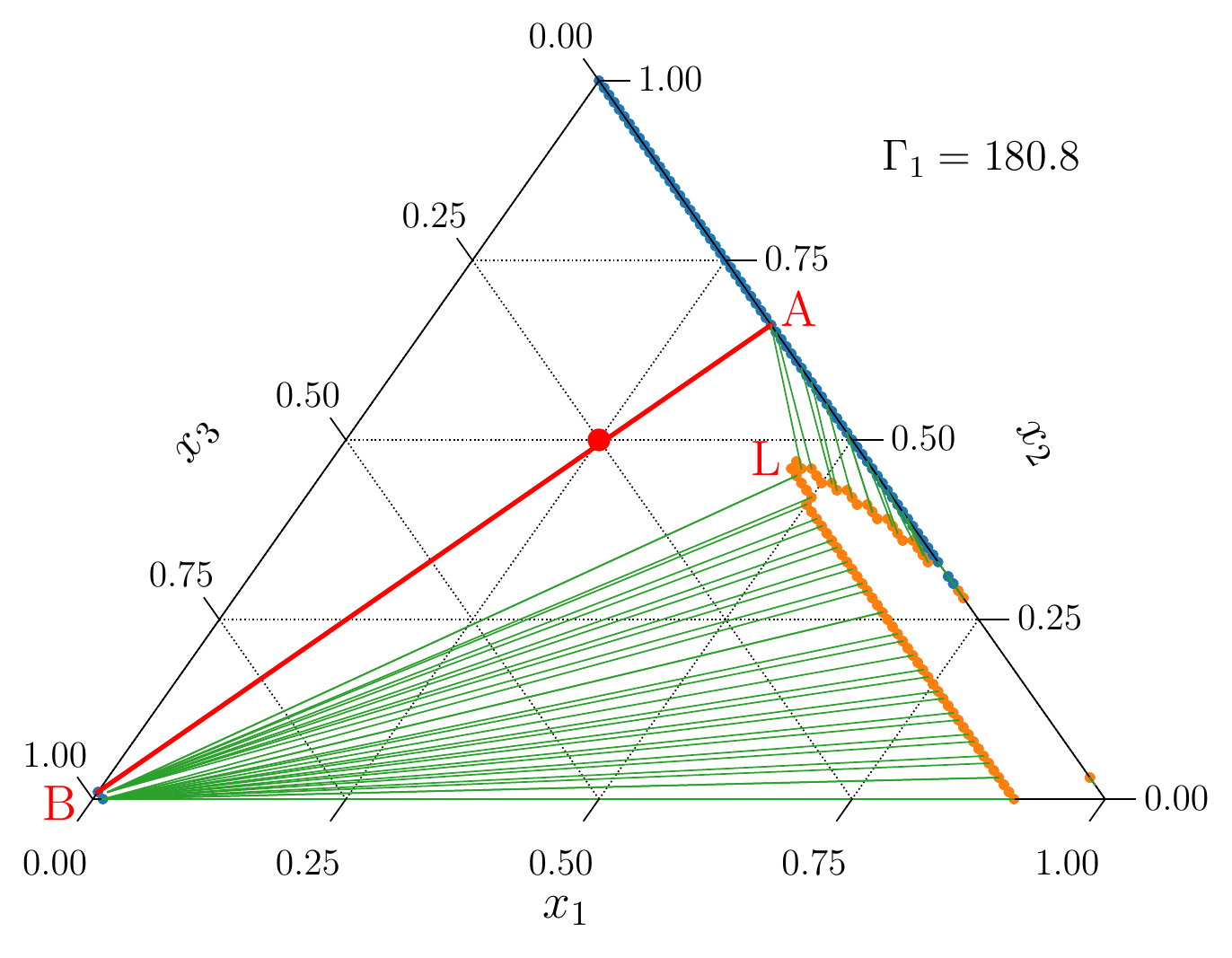}
\includegraphics[width=0.49\textwidth]{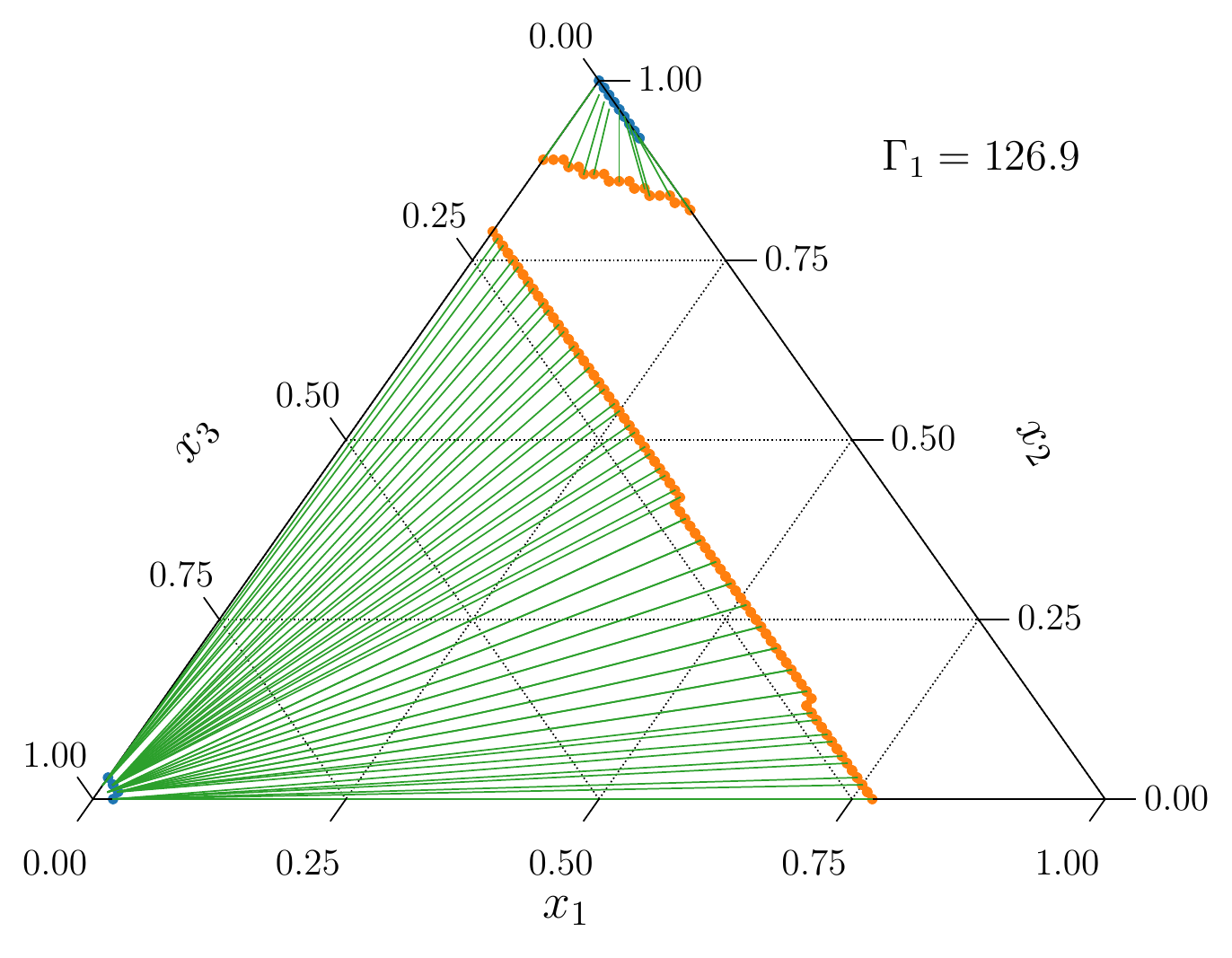}
\includegraphics[width=0.49\textwidth]{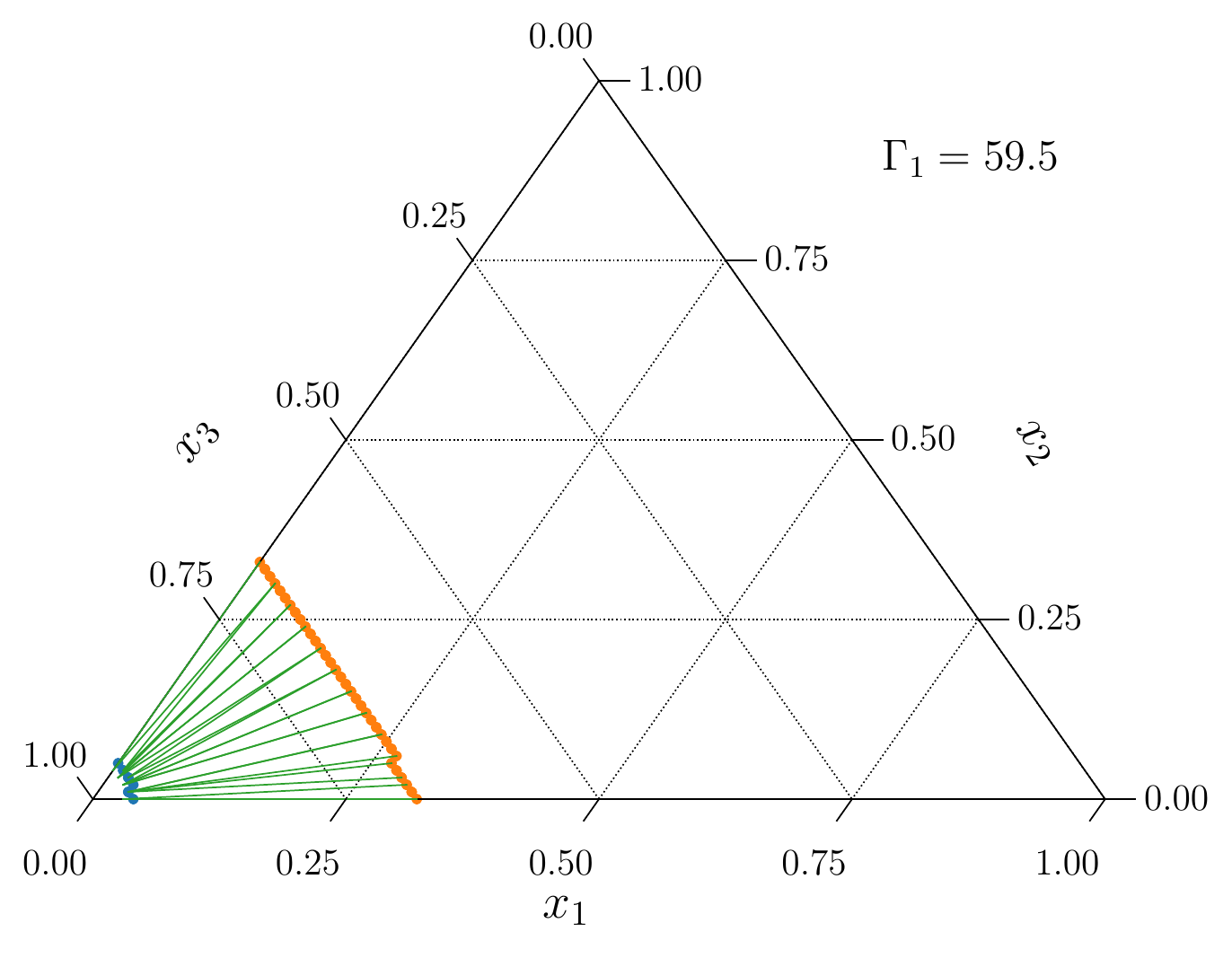}
\caption{\label{fig:phasediag2} 
Phase diagram for a three component mixture approximating rp-process ashes, with $Z_1=9$, $Z_2=12$, and $Z_3=22$. We show the liquidus (orange) and solidus (blue) at 4 different values of $\Gamma_1$, with tie-lines (green) linking liquid and solid points that are in equilibrium.  The one-component plasma melting points are at $\Gamma_1 = 178.6$, $110.6$ and $40.3$ for species 1,2, and 3 respectively. The solids that form are either mostly species 3 with small amounts of 1 and 2 (lower left), or a mixture of species 1 and 2 with negligible amounts of 3 (right side of the triangle). The red point at $(x_1,x_2,x_3)=(0.25,0.50,0.25)$ is a three-component approximation to the two similar compositions shown in Fig.~\ref{fig:Mckinven_comparison}. With small changes in the liquid composition near point L\added{=(0.46, 0.46, 0.08)}, solids A\added{=(0.32, 0.68, 0.00)} and B\added{=(0.01, 0.01, 0.98)} (marked in red on the plot) could alternately freeze out to form a polycrystalline crust that on average matches the composition of the rp-process ashes (red point). 
}
\end{figure*}

\begin{figure*}[t!]
\centering
\includegraphics[width=0.9\textwidth]{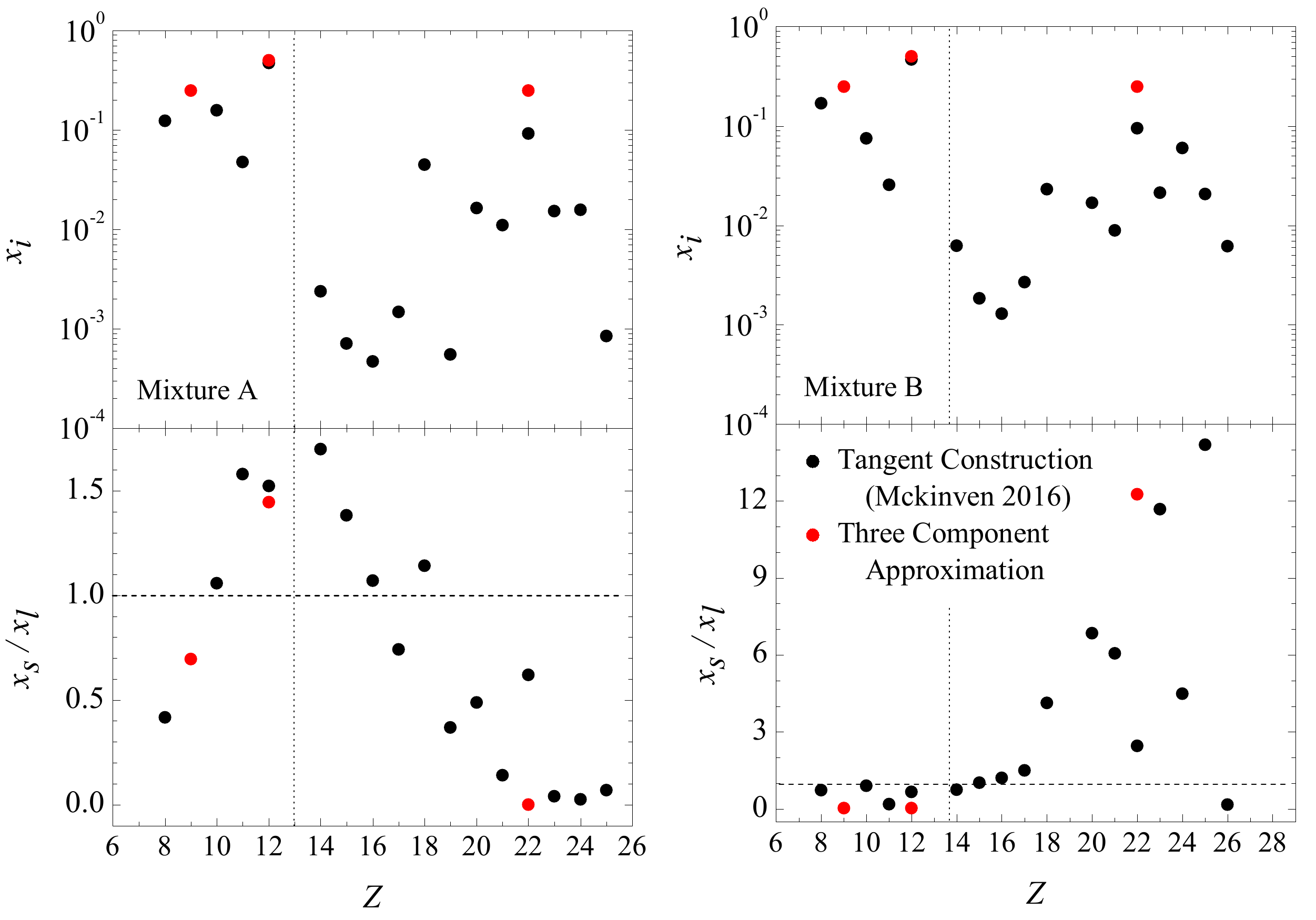}
\caption{\label{fig:Mckinven_comparison} 
As Fig.~\ref{fig:MD_comparison}, but for two mixtures from \cite{Mckinven:2016zkg} with similar compositions (top panels) but different phase separation behavior (bottom panels). We compare the full compositions (black) with a three-component mixture meant to approximate the total abundance and distribution of elements (red). The three-component mixture has $Z=9$, $12$, and $22$ with abundances $x_i=0.25$, $0.50$, and $0.25$ respectively. The $x_s/x_l$ ratios for the three-component mixture are derived from the $\Gamma_c/\Gamma_1=0.988$ phase diagram in Fig.~\ref{fig:phasediag2}, taking the liquid composition that is in equilibrium with the solids A and B marked in red on the plot.}  
\end{figure*}

The liquid and solid configurations were joined on one face to prepare a configuration with 55,296 nuclei, shown in the left panel of Fig. \ref{fig:MD}, the starting point of our simulation of phase separation. This half-solid/half-liquid configuration was evolved for $1,200 \times 10^6\ {\rm fm/c}$ ($48\times 10^6$ timesteps). The initial temperature was $0.042\ {\rm MeV}$ and was adjusted periodically in order to ensure that half of the simulation volume remained solid while half remained liquid, \added{because the melting temperature is not known precisely in advance (as in \cite{Horowitz:2007hx}). These temperature adjustments do not correspond to a physical process in the neutron star, rather, we seek to find the conditions for equilibrium solid-liquid coexistence, which is exactly the condition met by the ocean-crust interface in the neutron star.} To verify that the simulation had equal parts solid and liquid, we partitioned the simulation volume into 30 equal sized `slices' along the long axis every $10^7\ {\rm fm/c}$ and determined which slices were solid, liquid, or interfacial by inspection. 

The thermodynamic history is shown in the right panel of Figure~\ref{fig:MD}. The simulation equilibrated after $900 \times 10^6$ fm/c ($36 \times 10^6$ timesteps) \added{at a temperature of 0.044 MeV,} with no further changes in abundances or solid and liquid volume fractions. For each nuclear species, we show the ratio $x_s/x_l$ where $x_s$ ($x_l$) is the number fraction of nuclei of that species found in the solid (liquid) slices. Initially, the solid and the liquid have identical compositions ($x_s/x_l=1$) for all species. Within $100\times10^6$ fm/c ($4\times10^6$ timesteps), the compositions have begun to differentiate, nuclei with $Z=10$--$12$ enhanced ($x_s/x_l>1$) and other species depleted ($x_s/x_l<1$) in the solid.

We then extracted the final abundance ratios. Accurate identification of which particles in the simulation were in a solid configuration or in a liquid configuration was complicated by the appearance of a liquid subvolume within the crystal (this can be seen in Fig.~\ref{fig:MD}). To separate the liquid and solid abundances, we used the local bond order parameter, which quantifies the regularity of bond angles between nearest neighbors. This allows us quantify whether individual nuclei are in a lattice or an amorphous region more reliably than using inspection of domains. We use the method of \cite{Wang2005} (following \citealt{PhysRevB.28.784}) to calculate the bond order parameter,
\begin{equation}
Q_6 = \sqrt{ \frac{4 \pi}{2l+1} \sum_{m=-l}^{l} \left | \frac{1}{N_b} \sum_{\rm bonds} Y_{lm} ( \theta ( \mathbf{r} ) , \phi ( \mathbf{r} ) ) \right |^2 } ,
\end{equation}
for each nucleus. The spherical harmonics $Y_{lm}$ are calculated using the angles between pairs of nuclei $\theta ( \mathbf{r})$ and $\phi ( \mathbf{r} )$, averaged over nearest neighbors separated by vector $\mathbf{r}$ (number $N_b$ within $\left | \mathbf{r} \right | < 35$  fm). Nuclei are counted as solid if $Q_6 > 0.4$ and considered liquid if $Q_6 < 0.3$ ($Q_6=0.5$ for a zero-temperature bcc lattice). These cut-offs exclude approximately one third of our points, which are mostly near the interfaces or in the mixed region. 

Figure \ref{fig:MD_comparison} shows the final abundance ratios compared with the results from \cite{Mckinven:2016zkg}. The agreement is excellent, with the solid enhanced in $Z=10$--$12$ nuclei and depleted in heavier nuclei. This confirms the finding of \cite{Mckinven:2016zkg} that light mixtures phase separate differently than heavy mixtures. For comparison, we show in the left panel the molecular dynamics results of \cite{Horowitz:2007hx} for a heavy mixture. Heavy mixtures freeze a solid made up primarily of large $Z$ nuclei, with light elements staying in the liquid phase. Light mixtures on the other hand form a solid from the most abundant element. Depending on the composition, the solid can be lighter or heavier than the starting composition.

\section{Three-component phase separation }

Next, we calculate the phase diagram for a three-component mixture that approximates rp-process ashes, using it to explore the solid compositions that form. To model light, medium, and heavy elements we consider a three-component mixture with \replaced{$Z=9$, $Z=12$, and $Z=22$}{$Z_1=9$, $Z_2=12$, and $Z_3=22$, with fractional abundances of each species $x_i$ (defining composition $\vec{X}=(x_1, x_2, x_3)$)}. We use the double tangent construction (e.g.~\citealt{Gordon1968}) to identify unstable regions of the phase diagram, as described by \cite{Medin2010}. To compute the free energies of liquid and solid phases, we follow \cite{Medin2010}, specifically their equation (9) for the difference between the liquid and solid free energies for a one-component plasma (based on \citealt{Dubin1990} and \citealt{Dewitt2003}), and their equations (23) to (25) for the multicomponent liquid and solid (based on \citealt{Ogata1993}). 

\added{To briefy summarize our method, we begin with the free energies for solid and liquid mixtures,
\begin{equation}
\begin{split}
f_l^{ \rm{MCP} } & (\Gamma_1,x_1,\Gamma_2,x_2,\Gamma_3,x_3)  \simeq  \\ 
& \sum_{i=1}^3 x_i \left[ f_l^{ \rm{OCP} }(\Gamma_i) + \ln\left(x_i \frac{Z_i}{\langle Z \rangle}\right)\right]
\end{split}
\end{equation}
\noindent and
\begin{equation}
\begin{split}
f_s^{ \rm{MCP} } & (\Gamma_1,x_1,\Gamma_2,x_2,\Gamma_3,x_3)  \simeq  \\ 
& \sum_{i=1}^3 x_i \left[ f_s^{ \rm{OCP} }(\Gamma_i) + \ln\left(x_i \frac{Z_i}{\langle Z \rangle}\right)\right] \\
& + \delta f_s (\Gamma_1,x_1,\Gamma_2,x_2,\Gamma_3,x_3)
\end{split}
\end{equation}
with free energies for the one-component solid $f_s^{ \rm{OCP}}$ and liquid $f_l^{ \rm{OCP}}$, and a deviation from linear mixing for the solid $\delta f_s$ (from \citealt{Medin2010}).
}


\added{We use these free energies to find compositions that are unstable to phase separation.
To do this, we identify pairs of points on the minimum free energy surface which share a tangent plane (i.e. the double tangent construction).
Pairs of points found via the double tangent construction (i.e. pairs $\vec{X}_s$ and $\vec{X}_l$) correspond to solid compositions and liquid compositions that can coexist (i.e. phase separation). In Fig, \ref{fig:phasediag2} we plot all liquid compositions $\vec{X}_l$ found via the tangent construction in orange, with their corresponding solid composition $\vec{X}_s$ in blue. Connecting them in green are tie-lines, which span a metastable space. Any composition along this line will spontaneously separate into a solid with composition $\vec{X}_s$ and a liquid with composition $\vec{X}_l$. Thus, the region bounded by the orange curve consists of all stable liquid compositions, the region bounded by the blue points (for example along the $x_2$ axis and in the bottom left corner) consists of stable solid, while the remaining space in the phase diagram denotes an unstable region. In the unstable region, the mixture can lower its free energy by phase separating into a solid and liquid of the solidus and liquidus compositions at each end of the tie lines. For clarity, we show only tie lines connecting liquidus and solidus points, tie lines for solid-solid equilibria are not shown.
}

The results are shown in Figure \ref{fig:phasediag2} for 4 different values of $\Gamma_1 = Z_1^2e^2/a_ek_BT$. \deleted{In each case, we show the liquidus (orange), solidus (blue) and unstable regions (spanned by the green tie lines). In the unstable region, the mixture can lower its free energy by phase separating into a solid and liquid  of the solidus and liquidus compositions at each end of the tie lines. For clarity, we show only tie lines connecting liquidus and solidus points, tie lines for solid-solid equilibria are not shown.} For the free energy fits used here, a one-component plasma of species 1 would melt when $\Gamma_1=\Gamma_c=178.6$. However, a small amount of the heavy species 3 added to the mixture lowers the melting point, causing a liquid region to appear even for $\Gamma_1>\Gamma_c$ (top left of Fig.~\ref{fig:phasediag2}). As $\Gamma_1$ decreases, the liquidus sweeps across the phase diagram as progressively more mixtures are able to melt.

The striking feature of Figure \ref{fig:phasediag2} is that the solidus points are confined to the boundaries. Solids that form are either composed almost entirely of the heaviest element or are a mixture of species 1 and 2 and completely depleted in the heaviest element. This matches the behavior of the multicomponent mixtures in Figure \ref{fig:MD_comparison}, and is a result of the large charge ratio of the rp-process ashes. \cite{Ogata1993} showed for two-component plasmas that increasing the charge ratio between components significantly increases the free energy of the solid (see Fig.~3 of \citealt{Ogata1993}). This leads to an increasing preference for the mixture to phase separate into almost pure solids. As the charge ratio increases, the two-component phase diagrams progress from azeotropic/spindle type, with only a small difference between liquid and solid compositions, to eutectic type, with a large difference between liquid and solid (see Fig.~5 of \citealt{Ogata1993} and section 2.1 of \citealt{Medin2011}). We see a similar effect here for three-components. Previous studies of three-component plasmas focused on C-O-Ne mixtures for white dwarfs and found only small phase separations \citep{Ogata1993,Segretain1993,Hughto2012}.

\section{Implications for Accreted Crusts}

Both molecular dynamics with a full rp-process ash composition and a three-component approximation show that newly-forming crust in accreting neutron stars has a limited number of compositions available.
This means that for any given incoming mixture produced by the rp-process, there is not likely to be a solid with that composition that can form in equilibrium with the liquid at the base of the neutron star ocean. However, if multiple domains of distinct composition can form, then on average the ocean can reach a steady state and deposit out the incoming mixture. Figure \ref{fig:phasediag2} suggests a mechanism by which this could happen, since it shows that small changes in the liquid composition lead to dramatically different solid compositions. This can in fact be seen in the multicomponent results of \cite{Mckinven:2016zkg}. Figure \ref{fig:Mckinven_comparison} shows two compositions from that paper that are very similar yet phase separate differently (one in the same way as the light mixture and the other in the same way as the heavy mixture in Fig.~\ref{fig:MD_comparison}). 
The red point in the upper right panel of Figure \ref{fig:phasediag2} shows roughly where this mixture would lie in our three-component approximation.
Small changes in the ocean composition (point L\added{=($x_1$ = 0.46, $x_2$ = 0.46, $x_3$ = 0.08})) would cause the solid to alternate between two compositions, one \replaced{heavy}{light} (point A\added{=($x_1$ = 0.32, $x_2$ = 0.68, $x_3$ = 0.00)}) and one \replaced{light}{heavy} (point B\added{=($x_1$ = 0.01, $x_2$ = 0.01, $x_3$ = 0.98)}). The solid-liquid ratios between A-L and B-L 
are indicated in Figure \ref{fig:Mckinven_comparison}, and agree well with the multicomponent results. 

The idea of a polycrystalline crust composed of microscopic domains has been discussed before for a single composition (e.g.~\citealt{Kobyakov2015}). Here we see that formation of domains with different compositions is a natural outcome of the freezing of a many component plasma.
Support for this picture comes from molecular dynamics simulations of the solid phase \citep{HCB2009}, in which a solid initially created with one specific rp-process ash mixture phase separated into two solid components. 
In addition, when the composition was depleted in light elements, a single stable solid formed, with light nuclei occupying interstitial sites in the crystal lattice \citep{Horowitz:2009af}. 
Although our three-component approximation suggests that two solid phases form, in general many-component mixtures with large charge ratios may seek to form more than two crystal compositions. 

Our calculation does not specify the size of the compositional domains. They may take the form of microscopic grains with characteristic sizes determined by diffusion lengths if steady state deposition occurs. In that case, the ocean adopts a specific mixture (such as point L in Fig. \ref{fig:phasediag2}) and simultaneously freezes out multiple solids.
Diffusion could cause the domains to merge in the outer crust \citep{Mckinven:2016zkg}, although the diffusion rate drops exponentially and the tendency for separation into pure solid components increases at higher $\Gamma$ \citep{Ogata1993}.
Alternatively, the crust composition could switch on an astrophysical timescale associated with changing ocean composition,
as rp-process burning conditions change and as the ocean composition adjusts in response to losses to the crust. In that case, the domain size could be a significant fraction of the scale height ($\sim 10 \, \mathrm{m}$). Further work is needed to model compositional transport through the ocean, which will also have implications for X-ray bursts and superbursts. 
The fact that a three-component approximation is able to reproduce the phase separation of multicomponent mixtures suggests it may be promising to include in time-dependent simulations.

Several aspects of crustal physics should be revisited considering compositional domains. For example, new investigations into the transport properties of a polycrystalline crust are needed. While phase separation acts to increase the purity of the solid (so that individual domains should have a higher conductivity than the average composition), grain boundaries between microscopic grains may enhance electron scattering. Characterizing the crust with a single impurity parameter (e.g.~\citealt{Itoh1993}) may not be sufficient. The behavior of compositional domains as they undergo electron captures and other crust reactions will be important to study. In particular, the evolution of compositional domains through the crust may influence the size of the mass quadrupole associated with electron capture layers, a possible source of gravitational waves. Likewise, the partitioning of different species into domains may change pycnonuclear reaction rates \citep{Horowitz2008, Yakovlev2006}, a source of crustal heating. \added{In addition, a growing body of recent work is similarly emphasizing and investigating the importance of the transport and elastic properties of anistropic materials in the crust (such as in crystalline domains and nuclear pasta) \citep{Kobyakov2017,Kobyakov2018,Duriel2018}. }

\acknowledgments
This research was supported by Lilly Endowment, Inc., the Indiana University Pervasive Technology Institute, and the Indiana METACyt Initiative at Indiana University, and DOE grants DE-FG02-87ER40365 (Indiana University) and DE-SC0008808 (NUCLEI SciDAC Collaboration). M.~C.~is a CITA National Fellow. A.~C.~is supported by an NSERC Discovery grant. A.~C.~and M.~C.~are members of the Centre de Recherche en Astrophysique du Qu\'ebec, and thank Newcastle University for hospitality.


\end{document}